\documentclass[%
 reprint, amsmath,amssymb, aps,prb,floatfix]{revtex4-2}

\usepackage{graphicx}% Include figure files
\usepackage{dcolumn}% Align table columns on decimal point
\usepackage{bm}% bold math
\usepackage{hyperref}% add hypertext capabilities
\usepackage{xcolor}
\usepackage[mathlines]{lineno}% Enable numbering of text and display math
\usepackage{braket}
\usepackage{amsmath}
\begin{document}
%\preprint{APS/123-QED}
\title{Thermal Strain-Driven Periodic Ripples in Ultrathin Sb$_2$Te$_3$/Graphene Heterostructures and their Implications for Strain Engineering}
\author{Francisco Munoz}
\email{fvmunoz@gmail.com}
\affiliation{
Departamento de F\'isica, Facultad de Ciencias, Universidad de Chile \& CEDENNA, Santiago, Chile}

\author{Manuel Fuenzalida}
\affiliation{
Departamento de F\'isica, Facultad de Ciencias, Universidad de Chile \& CEDENNA, Santiago, Chile}

\author{Paula Mellado}
\affiliation{
School of Engineering and Sciences, 
	Universidad Adolfo Ib{\'a}{\~n}ez,
	Santiago, Chile}

\author{Caleb Z. Zerger}
\affiliation{Department of Physics, Stanford University, Stanford, CA 94305, United States}
    
\author{Hari C. Manoharan}
\affiliation{Department of Physics, Stanford University, Stanford, CA 94305, United States}

\author{Valentina Gallardo}
\affiliation{ Departamento de Ingeniería Mecánica, Universidad Técnica Federico Santa María, Avenida Espa\~na 1680, Valparaíso, Chile}

\author{Carolina Parra}
\email{carolina.parra@usm.cl}
\affiliation{ Departamento de Ingeniería Mecánica, Universidad Técnica Federico Santa María, Avenida Espa\~na 1680, Valparaíso, Chile}

%\date{\today}% It is always \today, today,
             %  but any date may be explicitly specified
\begin{abstract}

We report the observation of a highly regular, periodic ripple superstructure with a characteristic wavelength of approximately 8.7 nm in a van der Waals heterostructure composed of ultrathin Sb$_2$Te$_3$ epitaxially grown on monolayer graphene. We attribute this structural modulation to an extrinsic, thermally driven mechanism: upon post-growth cooling, the opposite signs of the thermal expansion coefficients of the bulk SiO$_2$ substrate and the graphene layer generate interfacial strain that drives the formation of ripples. In this ultrathin thickness regime, the Sb$_2$Te$_3$ film conforms to the buckled graphene morphology without delamination, because the interfacial van der Waals adhesion energy exceeds the elastic energy cost associated with bending. This substrate-induced templating mechanism is expected to be broadly applicable to other two-dimensional materials grown on graphene, provided that they are sufficiently thin to remain mechanically coupled below a critical bending thickness. Furthermore, based on density functional theory and effective low-energy model calculations, we predict that this structural corrugation leads to substantial modifications of the electronic band structure. The resulting periodic ripple potential suppresses the trivial hybridization gap in the planar configuration and transforms the spinless Dirac–Dirac resonance into a gapless helical metallic state with a dense, nontrivial spin texture.
\end{abstract}

\maketitle

\section{Introduction}
\label{sec:intro}

Straintronics,\cite{Miao2021} the control of electronic properties in two-dimensional materials via strain fields and folds, has become a promising way to create and control exotic electronic states. Of particular interest are periodic ripples, which induce a longer periodicity in one direction, creating an effective one-dimensional Moiré pattern\cite{Sinner2023}. 

Such ripples and other curvature effects \cite{WEN2023156156,LIU2024159792,RECKINGER2025164373,WU2016510} have been widely studied in graphene, including their formation and characterization\cite{Wang2013,meng2013,LIN2015207,SRUTRAKIC2016243}. They have an extrinsic origin, resulting from the small bending energy of graphene\cite{CHEN2023155979} and the negative thermal coefficient of graphene on a substrate with positive thermal expansion; hence, changes in temperature can create a large strain field \cite{Yoon2011}.

The effect of ripples in other Dirac materials, such as topological insulator (TI), has also been studied, however, to a much lesser extent. Okada \textit{et al.}\cite{Okada2012} showed a local modulation of the Dirac cone with the height profile in $\text{Bi}_2\text{Te}_3$. The ripples were induced by chemical disorder (defects), see also Ref.\cite{Shokri2016}. Salvato \textit{et al.} \cite{Salvato2022} found a very regular 1D ripple pattern in $\text{Bi}_2\text{Se}_3$, producing drastic local changes in the DOS and Fermi velocity close to the Dirac cone; the mechanism behind these ripples is unclear, as they ruled out many possibilities. Nevertheless, the mechanism is intrinsic, as the ripples have not been seen below a critical thickness. Rotation-induced Moiré patterns have been observed in ultrathin Sb$_2$Te$_3$ on graphene, attributed to the large lattice mismatch at different angles \cite{yin2022moire}. Strain-driven ripples at the edges have also been reported \cite{Lou_2018_nanolett}. All previous research shares a common point: the mechanism underlying the modulations is \textit{intrinsic}.

In this manuscript, we present an \textit{extrinsic} approach to creating ripples in 2D materials. In particular, we observed a ripple pattern with a wavelength of $d\approx8.7$~nm in ultrathin Sb$_2$Te$_3$ on graphene. The origin of the strain is thermal strain. If the Sb$_2$Te$_3$ layer is thin enough, the vdW energy can overcome the bending energy, resulting in highly regular ripples. While thermal strain in graphene often results in wrinkles, folds, and other irregular patterns, here, the Sb$_2$Te$_3$ layer stabilizes a quasi-periodic pattern. This extrinsic thermal approach should be readily applicable to other 2D materials, provided they are sufficiently thin. 

In Sec.~\ref {sec:experimental}, we show the experimental realization of a rippled vdW heterostructure of Sb$_2$Te$_3$ on single-layer graphene (SLG). In Sec.~\ref{sec:theo_origin} we explain the ripple pattern found as an extrinsic thermal effect, which is strongly supported by energy considerations in Appendix~\ref{app:simpleEnergy}. Then, in Sec.~\ref{sec:electronic}, by means of density functional calculations and a theoretical model, we explore the expected effects on the electronic structure due to the ripples on the Sb$_2$Te$_3$-graphene heterostructure.  Finally, we draw our conclusions.

%%%%%%%%%%%%%%%%

\section{1QL  Sb\textsubscript{2}Te\textsubscript{3} nanoplatelet grown on single-layer graphene}
\label{sec:experimental}

\begin{figure}
    \centering
    \includegraphics[width=\columnwidth]{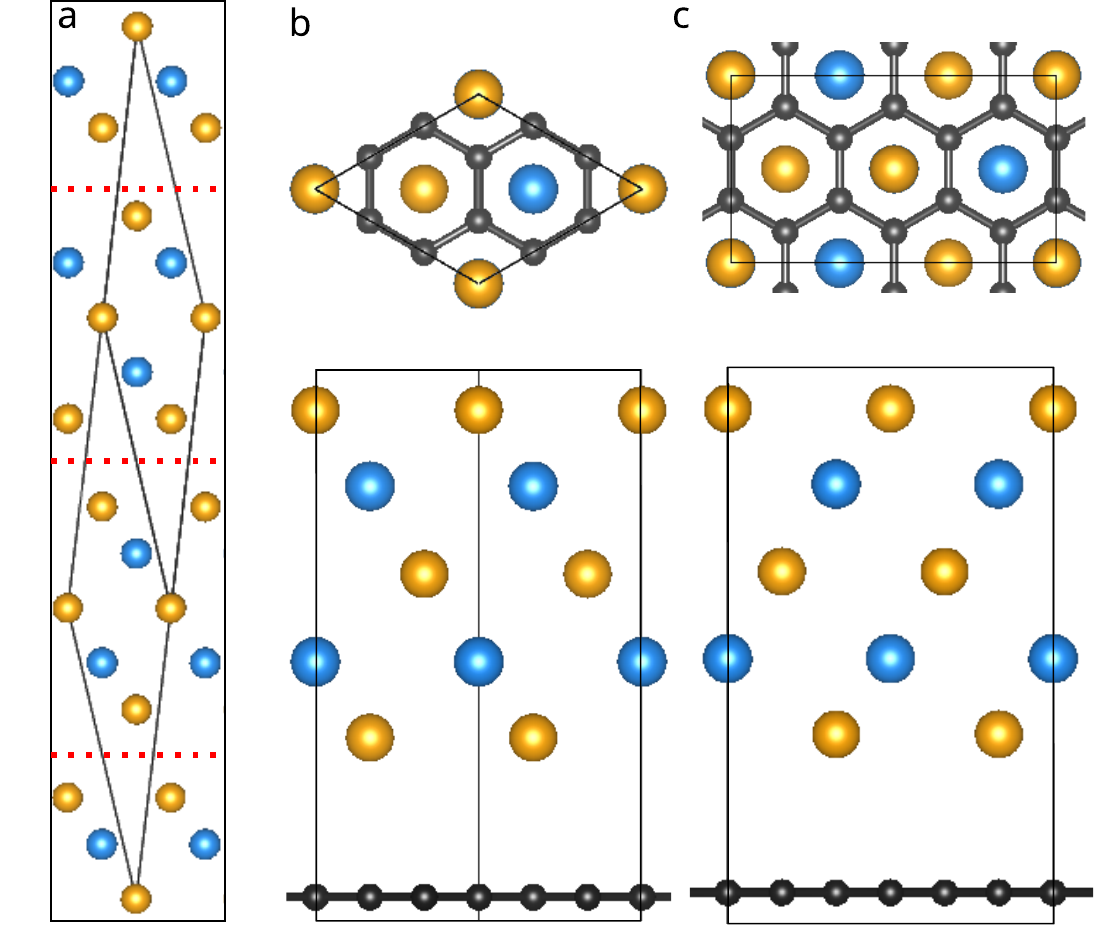}
    \caption{(a) Crystal structure of Sb$_2$Te$_3$. The rhombohedral primitive cell is denoted by the inner black lines, and the quintuple layers are separated by dotted red lines. Each QL is nearly 1 nm wide. Along the article Te, Sb, C atoms are colored orange, cyan and black, respectively. (b) Top and lateral view of the primitive cell of 1 QL of Sb$_2$Te$_3$ on graphene (vacuum space is trimmed). (c) orthogonal cell of the heterostructure used to calculate buckling.}
    \label{fig:geometries}
\end{figure}

$\text{Sb}_2\text{Te}_3$ is a tetradymite semiconductor that has been identified as a three-dimensional strong topological insulator (TI) and exhibits a rhombohedral crystal structure with the space group D$^5$$_3$$_d$ or R3m \cite{Zhang2009}. This material has a layered structure, with each layer consisting of a triangular lattice of a single atomic species, either Sb or Te. Five such layers constitute one quintuple layer (QL) unit Fig.~\ref{fig:geometries}a shows the structure of the material. Covalent bonding predominates within each QL, while the interaction between QLs is primarily van der Waals in nature. As a result, $\text{Sb}_2\text{Te}_3$ is typically manipulated in units of QLs, and the analysis here focuses on a single QL. 

\begin{figure*}[ht]
\centering
\includegraphics[width=\textwidth]{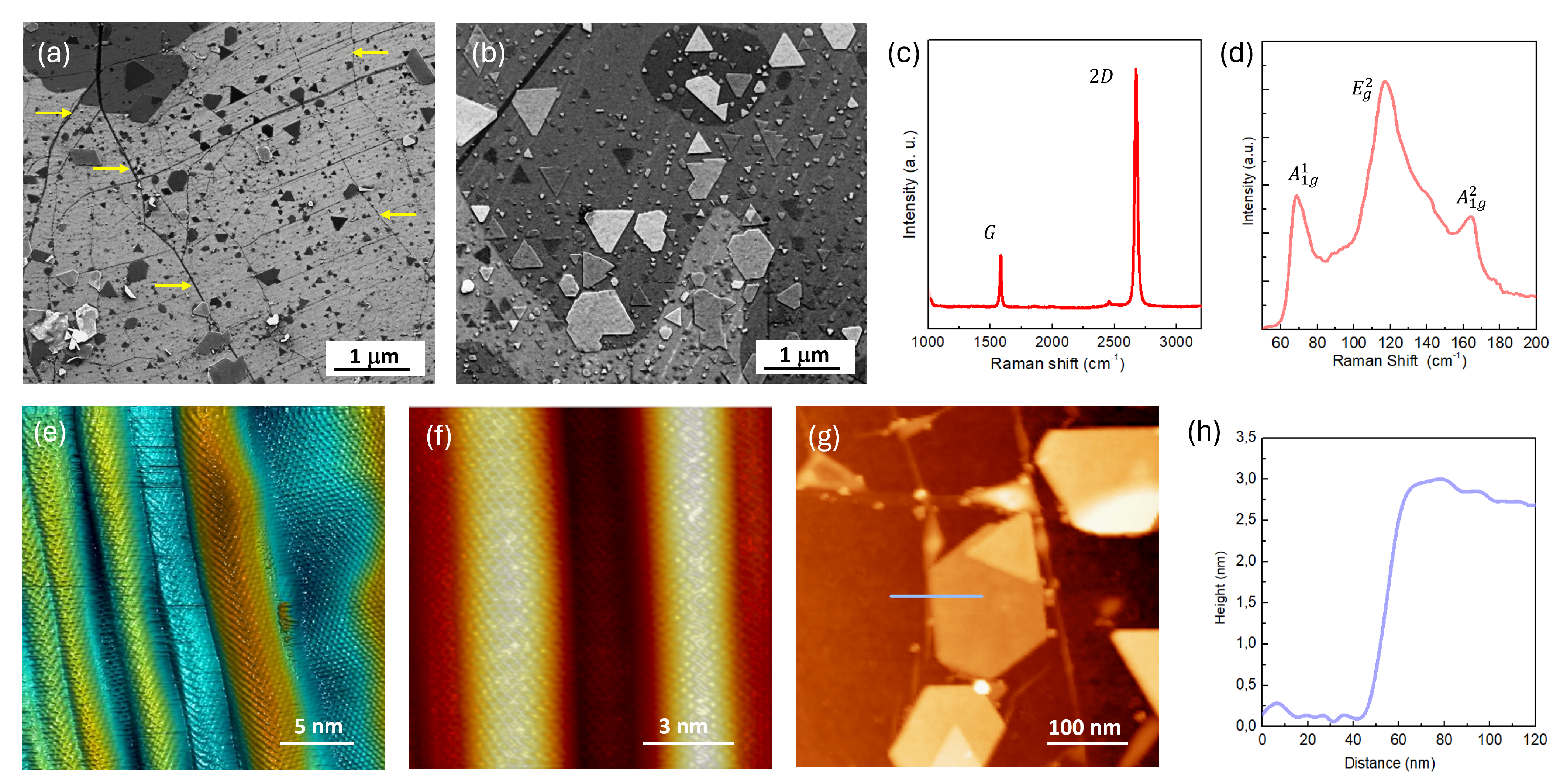}
\caption{(a) and (b) SEM micrographs of Sb$_2$Te$_3$ nanoplatelets grown on single-layer graphene (SLG)/SiO$_2$. Yellow arrows indicate pre-existing wrinkles in the SLG substrate. The triangular morphology of the platelets reveals two primary orientations rotated by $180^\circ$, identifying the presence of twinning defects discussed in the sec.~\ref{sec:electronic}. (c) Representative Raman spectra of SLG/SiO\textsubscript{2} (d) Representative Raman spectra of Sb\textsubscript{2}Te\textsubscript{3} nanoplatelets grown on SLG/SiO\textsubscript{2}, (e) and (f) STM topography of uniaxial ripples found in CVD SLG transfer onto SiO\textsubscript{2}, after the cooling-down process. (g) AFM image of Sb\textsubscript{2}Te\textsubscript{3} nanoplatelets grown on graphene/SiO\textsubscript{2} including a height profile for a 3QL thick Sb\textsubscript{2}Te\textsubscript{3} nanoplatelet (h)}
\label{fig:STM}
\end{figure*}

\begin{figure*}[ht]
\centering
\includegraphics[width=\textwidth]{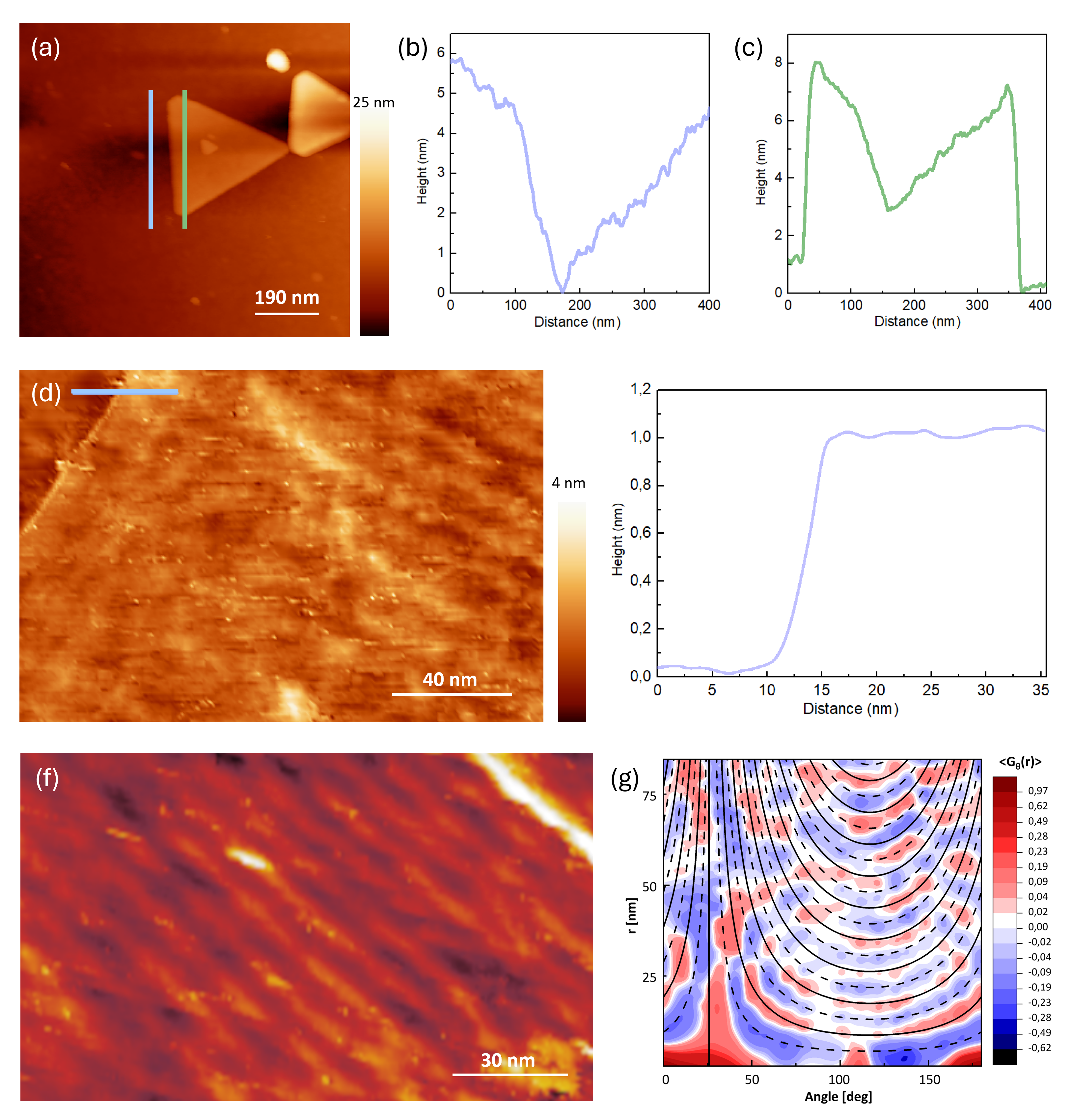}
\caption{(a) AFM image of  Sb\textsubscript{2}Te\textsubscript{3} NP grown on single-layer graphene, profiles in (b) and (c) show how the topography of the substrate (blue) is reproduce in the NP (green), (d) STM topography of a 1QL  Sb\textsubscript{2}Te\textsubscript{3} nanoplatelet grown on SLG with the corresponding heigth profile (e), (f) STM topography of the modulated pattern found in the surface of the 1QL NP. (b) Angle-dependent correlation function $<$G\textsubscript{$\theta$}(r)$>$ for the stripes in (f). The color scale on the right-hand side plots represents the value of  $<$G\textsubscript{$\theta$}(r)$>$. This quantity is plotted as a function of $|r|$ (vertical axis) and the angle $\theta$ with the horizontal axis. The angle-dependent correlation functions reveal clear arcs of intensity, associated with the stripe-like modulation on the surface of Sb\textsubscript{2}Te\textsubscript{3} nanoplatelet. Solid and dashed lines represent the best fits to $Nd/\cos[(\alpha–90^\circ)–\theta]$ and $(2N–1)w/\cos[(\alpha–90^\circ)–\theta]$ for the local maxima and minima, respectively, as described in the main text.}
\label{fig:STM2}
\end{figure*}

Thin Sb$_2$Te$_3$ nanoplatelets (NPs) were grown on SLG (Fig. \ref{fig:STM}a,b), forming a nano-heterostructure. The nanoplatelets exhibit the typical triangular pattern and orientation imposed by the graphene substrate \cite{Gallardo}. The TI synthesis parameters were tuned to obtain nanostructures thinner than 30 nm, \textit{i.e.} below the 3D to 2D crossover limit\cite{Zhang2010}, to \textit{(i)} promote the hybridization between top and bottom surface states, and their interaction with the SLG Dirac cone, and \textit{(ii)} to keep the Sb$_2$Te$_3$ able to be affected by external strain. Sb$_2$Te$_3$ NPs were synthesized by a catalyst-free vapor transport and deposition process.  The resulting nano-heterostructures were characterized by SEM ((Fig.\ref{fig:STM}a,b), Raman spectroscopy (Fig. \ref{fig:STM}d), AFM (Fig. \ref{fig:STM}g, Fig. \ref{fig:STM2}a), and low-temperature scanning tunneling microscopy LT-STM (Fig. \ref{fig:STM2}d,f), to obtain information on their morphological, compositional and structural properties, see Appendix\ref{sec:app-experimental} for additional details. Fig.~\ref{fig:geometries}b-c shows the expected atomic geometry of the heterostructure, with almost no strain, as calculated using density functional theory (see next section).

Fig.\ref{fig:STM2}f shows a buckling pattern observed on the surface of a 1QL Sb$_2$Te$_3$ NP grown on single-layer graphene (Fig.\ref{fig:STM2}d). The angle-dependent spatial correlation function $\langle G_\theta(r)\rangle$ (Fig.\ref{fig:STM2}g) was calculated for the topographic image shown in Fig.\ref{fig:STM2}f. $\langle G_\theta(r)\rangle$ reflects the spatial pattern observed in the topographic image, which shows a particular spatial pattern with a characteristic length scale \cite{Giraldo2015}.  In Fig.\ref{fig:STM2}g, arcs of intensity that repeat with a fixed periodicity are marked; these imply a pattern of phase separation of low and high topographic regions over large length scales. These intensity arcs are consistent with the real-space phase separation of a topographic striped pattern. The functional form of these arcs for a system with stripes separated by a distance $d\approx8.7$~nm and running along an angle $\alpha$ from the horizontal, measured at an angle $\theta$, is Nd/cos(($\alpha$-90)-$\theta$). 

Due to thermal cycling, buckling of graphene can be triggered by compressive strain \cite{Yoon2011}. Under such strain, graphene buckles out of plane, forming wrinkles, ripples, or bubbles. Fig.\ref{fig:STM}e and Fig.\ref{fig:STM}f show STM topographies of the ripples found in the SLG transferred to SiO\textsubscript{2}, after post-growth thermal treatment. The nano-heterostructure sample presented here undergoes a thermal treatment prior LT-STM measurements as well. A thin TI nanoplatelet grown on this strained graphene substrate has been found to present spatial height modulation, forming corrugated ripples, as seen in Fig.\ref{fig:STM2}a. Although the height modulation of the observed buckling pattern in the NP is around 0.6 nm, the mean surface roughness is around 0.18 nm, a value similar to the one found in buckled graphene superlattices \cite{Mao2020}.


\section{Origin of the ripples in samples}
\label{sec:theo_origin}

From DFT calculations, we obtained a lattice constant of $a_{g}=2.456$~\AA{} for the graphene monolayer, and $a_{1QL}=4.269$~\AA{} for a single quintuple layer (QL) of Sb$_2$Te$_3$. A $\sqrt{3}\times\sqrt{3}$ supercell of graphene (\textit{e.g.} 6 atoms, a full hexagon) has a lattice constant of 4.254 \AA, with just 0.35\% of mismatch with the Sb$_2$Te$_3$ monolayer, see Fig.~\ref{fig:geometries}b-c. 

To determine the physical origin of the observed ripples, we first evaluated whether the corrugation arises from an intrinsic instability of the freestanding $\text{Sb}_2\text{Te}_3$/graphene heterostructure or from the formation of distinct material phases. Our energetic analysis (detailed in Appendix~\ref{app:simpleEnergy}) yields three key findings. First, the van der Waals binding energy between the layers ($E_{bind} \approx 0.23$ eV/cell) is significantly larger than the estimated bending energy of the rippled stack (estimation of an upper bound, $E_{bend} \approx 0.20$ eV/stripe, a stripe repeats $\sim12$ times the cell of Fig.~\ref{fig:geometries}c), ensuring that the two materials remain coupled rather than delaminating. Second, the energy penalties for relative sliding and intrinsic strain accumulation are negligible ($<0.01$ eV), ruling out stress-induced buckling from lattice mismatch. Third, the formation of alternating Sb$_x$Te$_{1-x}$ phases is energetically unfavorable due to high phase-boundary costs. Furthermore, molecular dynamics simulations of the freestanding system show no spontaneous stripe formation. Consequently, the ripples cannot be intrinsic to the heterostructure itself and must be driven by external factors.

Having ruled out intrinsic mechanisms, we proposed that the observed ripples originate from a pre-existing buckled structure in the graphene substrate. The formation of this structure can be understood as a two-step process: substrate preparation followed by growth. First, during the wet transfer process, the graphene film is deposited from solution onto the SiO$_2$ substrate. As the meniscus drains, directional capillary forces can create an initial, anisotropic set of wrinkles aligned along a preferential axis~\cite{Calado2012}, imposing a one-dimensional structural template on the graphene before any additional processing. Next, the Graphene/SiO$_2$ substrate is heated for Sb$_2$Te$_3$ growth and subsequently cooled. Due to the mismatch in thermal expansion coefficients between graphene and SiO$_2$~\cite{Yoon2011}, significant thermal strain develops. Because graphene has a negative thermal expansion coefficient (expanding upon cooling) while the substrate contracts, the sheet experiences compressive strain. To relieve this, the graphene buckles along the previously formed wrinkles, which serve as lines of weakness, amplifying a highly periodic ripple pattern. The Sb$_2$Te$_3$ monolayer grows conformally on this pre-rippled graphene template.  This model also explains why the effect is not observed for thicker films: as energetic considerations suggest, the bending energy of a stack thicker than 1QL would be too high for the available thermal strain energy to overcome, making buckling energetically unfavorable. For thicker NPs (Fig \ref{fig:STM2}a), larger height variations in the substrate topography (height profile in blue in Fig \ref{fig:STM2}b) are reproduced in nanoplatelet topography (height profile in green in Fig \ref{fig:STM2}c).

Unlike the randomized or highly localized wrinkles typically observed when bare SLG undergoes thermal buckling, the measured Sb$_2$Te$_3$/graphene heterostructure exhibits highly regular, periodic ripples with a well-defined wavelength. This structural coherence is mechanically enforced by the addition of the Sb$_2$Te$_3$ layer. The elastic bending energy of the system scales quadratically with its local curvature, $\kappa$ ($E_{bend} \propto \kappa^2$). Because Sb$_2$Te$_3$ has a significantly larger thickness and bending modulus compared to bare graphene, forming sharp, irregular folds would incur a massive energetic penalty. Therefore, to relieve the substrate-induced compressive strain most efficiently, the heterostructure avoids chaotic fluctuations and instead distributes the deformation into a smooth, continuous, and energetically favorable periodic ripple.

The thermal mechanism described above, driven by the negative thermal expansion coefficient of SLG, is broadly applicable to other vdW heterostructures, such as transition metal dichalcogenides and other two-dimensional materials grown on graphene. Because the elastic bending energy scales quadratically with thickness ($E_{bend}\propto t^2$), there exists a critical thickness threshold where this bending penalty remains smaller than the interfacial van der Waals binding energy ($E_{bend}<E_{bind}$). Below this critical thickness, the minimal energy conformation is for the 2D overlayer to bend conformally with the underlying graphene. Consequently, the top 2D material acts as a mechanical stabilizer, regularizing the substrate's thermal stress into a highly periodic ripple pattern.

\section{Prediction of Ripple-Induced Helical Metal States}
\label{sec:electronic}

To isolate the electronic effects induced by the structural corrugation, it is instructive to first summarize the properties of the ideal, flat heterostructure (a detailed analysis is provided in App.~\ref{sec:flat}). In its freestanding form, a single quintuple layer (1 QL) of $\text{Sb}_2\text{Te}_3$ exhibits a trivial bandgap of $0.45$ eV, which arises from the strong spatial hybridization between the top and bottom surface states at this ultrathin limit. However, when this 1 QL film is supported on SLG, the topological insulator surface states and the graphene's Dirac cone undergo a strong hybridization at the $\mathrm{\Gamma}$ point. This interaction yields a Dirac-Dirac resonance, splitting them into two pairs of massive Dirac cones located directly at the Fermi level with respective gaps of $10$ meV and $40$ meV. Notably, this strong resonant coupling is highly sensitive to quantum confinement and is significantly diminished in thicker samples where the top and bottom surface states decouple\cite{kiemle2022gate,song2018spin}.

\begin{figure}
    \centering
    \includegraphics[width=\columnwidth]{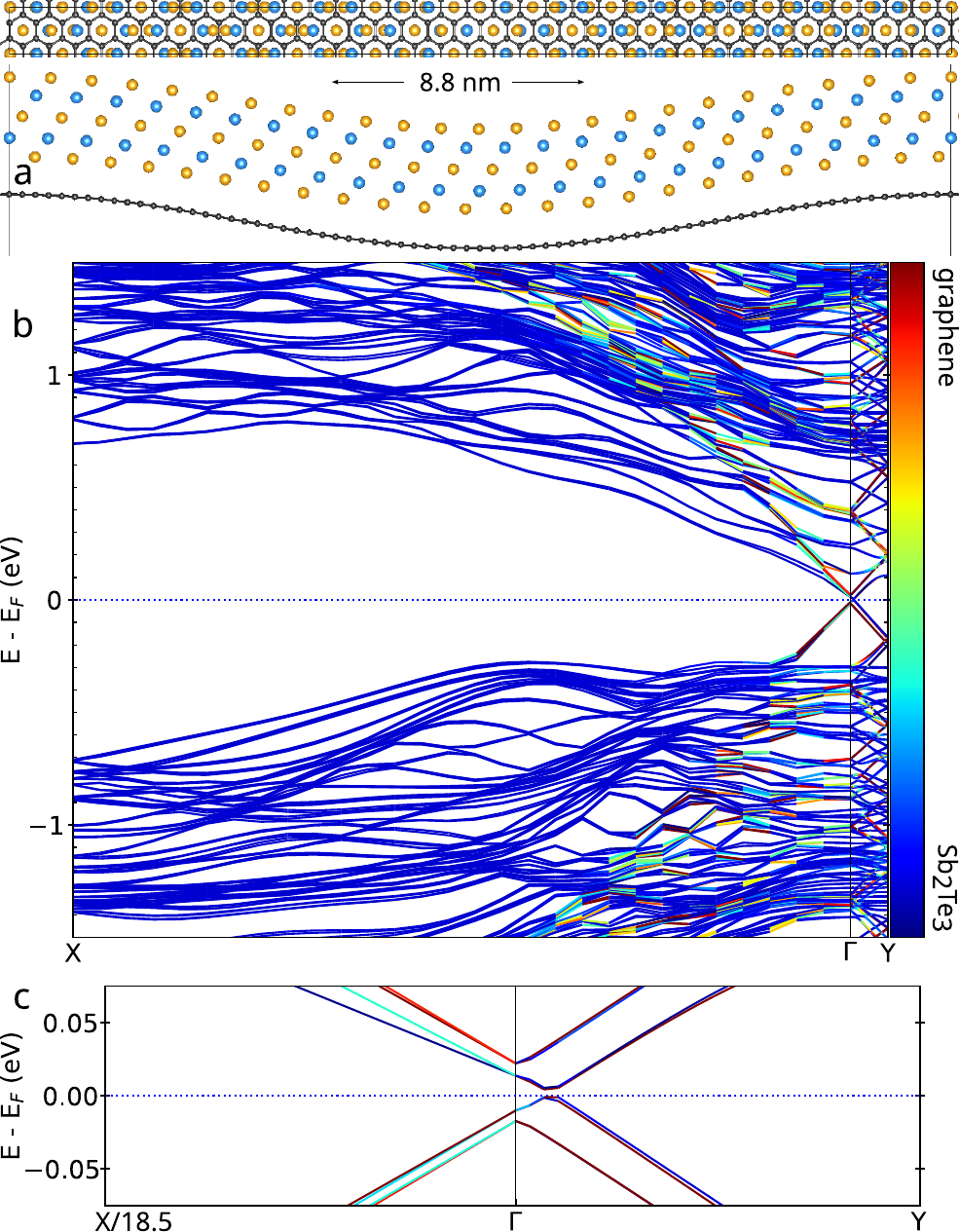}
    \caption{(a) Top and lateral views of the supercell used to model the electronic structure of the ripples. (b) Band structure of the rippled supercell; X and Y correspond to directions perpendicular to and along the ripples, respectively. (c) Zoom of the low-energy bands near the Fermi level.}
    \label{fig:buckled}
\end{figure}

To characterize the electronic influence of the buckling, we modeled the simplest periodic ripple case, excluding secondary effects such as shear stress. For this purpose, we employed the orthogonal supercell shown in Fig.~\ref{fig:buckled}a. Its short axis corresponds to the pristine unit cell width, while the long axis is repeated to match the experimentally observed ripple wavelength. Although not presented here, calculations for six other wavelengths yielded qualitatively similar results. In the structural relaxation, the out-of-plane ($z$) positions of the graphene atoms were fixed to the cosine profile of the STM results, while all other degrees of freedom were allowed to relax. The supercell lattice vector along the ripple direction was shortened to conserve the total arc length of the graphene sheet relative to its flat state, thereby mimicking the physical relaxation of compressive strain (see Appendix~\ref{sec:computational} for details).

The resulting electronic structure is shown in Fig.~\ref{fig:buckled}b. On a large energy scale, it resembles the flat unit cell but folded into the smaller mini-Brillouin zone defined by the long-period superlattice. However, the low-energy physics (Fig.~\ref{fig:buckled}c) reveals two striking phenomena distinct from the flat case: \textit{(i)} the Dirac cone is shifted away from the high-symmetry $\Gamma$ point, and \textit{(ii)} this shifted Dirac cone becomes effectively gapless ($<1$ meV).

The momentum shift of the Dirac cone is expected. The local lattice deformation (compression and elongation) induced by buckling creates a pseudo-vector potential $\mathbf{A}$ that acts as a gauge field, shifting the Dirac point from $\mathbf{K}$ to $\mathbf{K} - \mathbf{A}$ without opening a gap~\cite{banerjee2020} (note that $\mathbf{K}$ is folded to $\Gamma$ in the supercell). Although the varying strain field can theoretically induce a pseudo-Landau level structure, such quantization is not observed here, likely because the strain magnitude is too small or due to hybridization effects. Crucially, unlike the bare case, the graphene cone here remains hybridized with the Sb$_2$Te$_3$ surface states. The fact that the gap closes (restoring metallicity) suggests that the symmetry breaking induced by the ripple disrupts the specific hybridization conditions that opened the $\sim40$ meV gap in the flat heterostructure.

It is important to note that this buckling mechanism fundamentally differs from electronic instabilities like Charge Density Waves (CDW);%. In a CDW, a lattice distortion occurs spontaneously to lower the total electronic energy by opening a gap at the Fermi level. In our case, 
the buckling increases the elastic energy of the Sb$_2$Te$_3$ layer without a compensating electronic gain. Quite the opposite: %our DFT calculations show that 
the system transitions from a gapped (flat) state to a gapless (rippled) state, increasing the population of states at the Fermi level. This confirms that the ripples are driven not by electronic energy minimization, but by the extrinsic release of thermal strain from the substrate.

%\subsection{Emergence of a Helical Metal}

%The transition to a gapless state in the presence of ripples raises a fundamental question about the topological character of the quasiparticles. In the flat heterostructure, hybridization opens a gap, effectively destroying the single Dirac cone. 

One might expect that the additional disorder from the ripples would result in a trivial metallic state without a clear spin character. To understand the interplay between the ripples, acting as a moir\'e modulation, and the spin-orbit coupling, we developed an effective ladder model (detailed derivation and formalism in Appendix \ref{app:effective}). The model reveals that while the ripples modifies the simple band structure, it promotes a spin-momentum locking. Instead of a clean Rashba splitting, the moiré potential generates a dense manifold of minibands (see Appendix Fig.~\ref{fig:model_1}), where the helicity is distributed across the entire low-energy spectrum \cite{mellado2026moire}. This suggests that the system behaves as a ``Helical Metal'': a state where the density of states at the Fermi level is composed entirely of states with strong spin-momentum locking. Helical states have been observed in moiré heterostructures featuring proximity induced SOC \cite{takagaki2016manipulation,khokhriakov2020gate,kiemle2022gate,song2018spin,li2012topological,hoque2024room,tanabe2025high,Zollner_2025}. In the case of Sb$_2$Te$_3$ , combined Angle-Resolved Photoemission Spectroscopy (ARPES) and electrical detection experiments provide strong experimental evidence for the existence and manipulability of the helical surface states where electron spin is locked perpendicular to momentum due to strong spin-orbit coupling and time-reversal symmetry \cite{li2016electrical,takagaki2016manipulation,takagaki2012robust,zollner2021heterostructures,yin2022moire,zhao2024layer,locatelli2022magnetotransport,manchon2015new}. 

% We verified this prediction by calculating the spin texture of the rippled supercell with DFT. Fig. \ref{fig:spin_texture} displays the spectral weight projected onto the Cartesian spin components ($S_x, S_y, S_z$). 

\begin{figure}
    \centering
    \includegraphics[width=\linewidth]{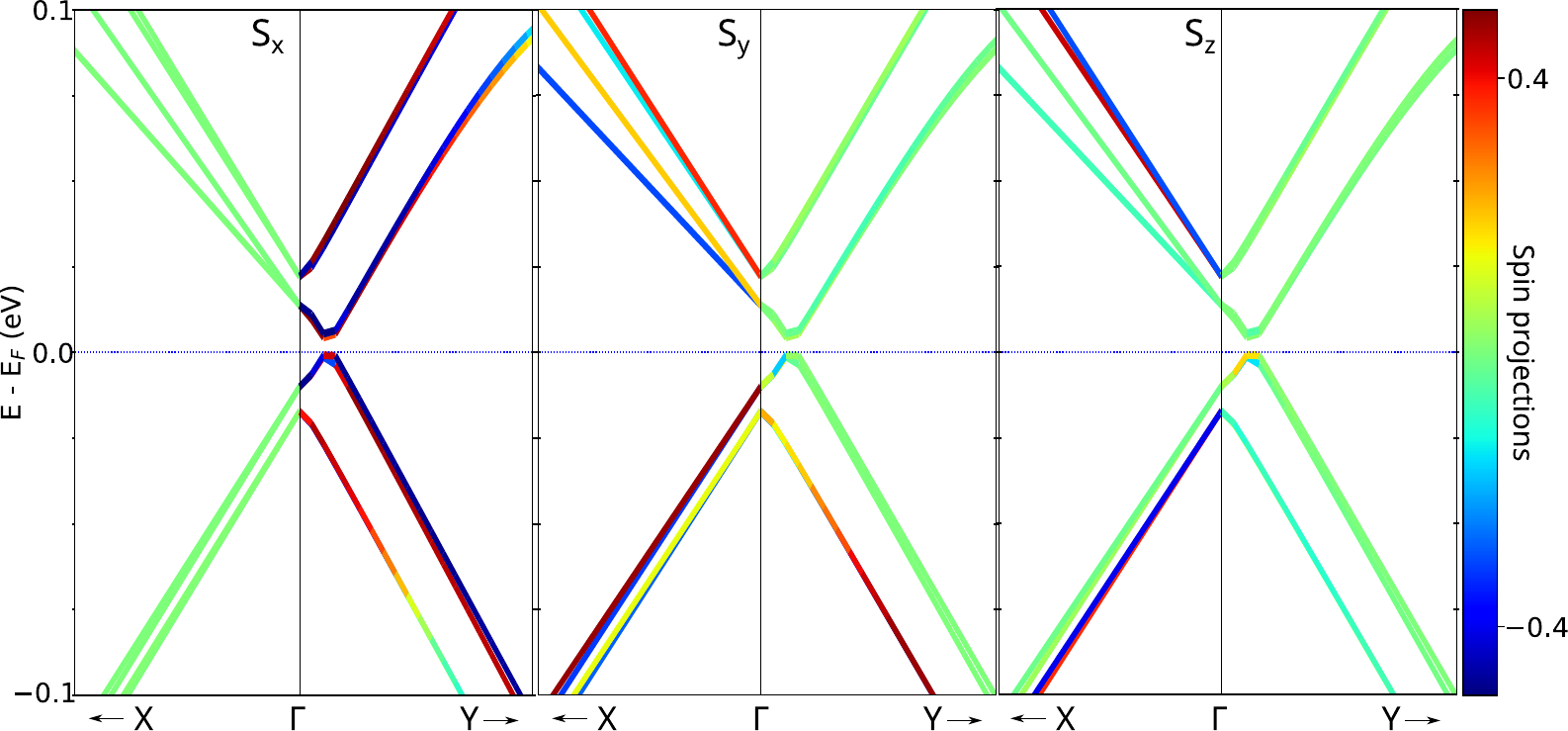}
    \caption{Spin-texture close to the Dirac cone of the system with ripples. The spin projection axes ($S_x,S_y,S_z$) coincides with the Cartesian axes.}
    \label{fig:spin_texture}
\end{figure}

In the flat 1QL limit at this energy scale, the bands exhibit negligible spin texture due to the hybridization gap. In strong contrast, the rippled heterostructure reveals a vibrant spin texture. As such, in principle, induce a pseudo-Landau-level structure, such quantization is not observed here, likely because the strain magnitude is too small or because those along $\Gamma-X$ are polarized in $S_y$ (center panel). This orthogonal locking ($k_y \perp S_x, k_x \perp S_y$) is characteristic of Rashba-like interactions, yet the texture is far more complex than a standard Rashba model. The polarization is not confined to a single pair of bands but is distributed across multiple crossing branches, confirming the existence of the helical manifold predicted by our model. The out-of-plane component $S_z$ is negligible along $\Gamma-Y$, but it can be as large 0.5 along $\Gamma-X$, in a strong departure from the Rashba model. While the ripples ``smear'' the electronic states in momentum space (\textit{i.e.} minibands), it creates a well-defined non-trivial spin-texture, which could be a rich platform for spin-dependent transport \cite{Bordoloi2024} despite the lack of a protected topological gap. It is worth noting that the structure of the nanoplatelets observed in the SEM images (Fig.~\ref{fig:STM}) is characterized by $ 180^\circ$-rotated triangular domains, indicating the presence of twinning defects, a common result of the weak vdW coupling between the Sb$_2$Te$_3$ quintuple layers and the graphene substrate\cite{Jung2017}. From a transport perspective, these twin boundaries represent regions of spin-texture discontinuity. As the helical surface states encounter a domain wall, the requirement for spin realignment introduces a significant scattering potential\cite{laliena}. Furthermore, in the context of the moir\'e ladder model, these boundaries may act as phase-slips that fragment the predicted miniband structure\cite{mellado2026moire}. For the emergent helical metal to remain robust, the growth parameters must be tuned to promote single-domain coalescence.

\section{Conclusions}

In summary, we have presented a comprehensive structural and theoretical investigation of periodic ripple formation in ultrathin $\text{Sb}_2\text{Te}_3$/graphene van der Waals heterostructures. Our low-temperature STM measurements reveal a distinct, highly regular buckling pattern, which we demonstrate is not intrinsic, but rather an extrinsic effect driven by the thermal expansion mismatch between the graphene layer and the bulk $\text{SiO}_2$ substrate during the post-growth cooling phase. We establish that at the ultimate two-dimensional limit—a single quintuple layer of $\text{Sb}_2\text{Te}_3$—the interfacial van der Waals binding energy effectively overrides the elastic bending energy penalty, ensuring that the topological insulator conforms to the buckled substrate rather than delaminating or flattening the ripples. Crucially, the presence of the relatively stiff $\text{Sb}_2\text{Te}_3$ layer acts as a mechanical regularizer, preventing chaotic crumpling and stabilizing the substrate's thermal stress into a well-defined, periodic ripple pattern. Because this mechanism relies on macroscopic thermodynamic forces and continuum mechanics, it presents a highly generalizable pathway for deterministically engineering strain and curvature in other substrate-supported 2D materials, provided they remain below their critical bending thickness. 

Based in the experimentally found ripples pattern, by means of DFT calculations and theoretical models, we explored how the periodic rippling should alters the electronic landscape of the heterostructure. We predict that the structural corrugation breaks the hybridization symmetry, closing the trivial gap associated with the flat 1 QL limit and triggering the emergence of a gapless ``Helical Metal'' state. Rather than experiencing a simple Rashba splitting, the spinless Dirac-Dirac resonance of the flat system develops a dense, non-trivial spin polarization distributed across a manifold of moir\'e minibands. This geometry-induced entanglement of spin, sublattice, and leg degrees of freedom creates an emergent helicity—featuring orthogonal spin-momentum locking along one axis and a strong out-of-plane component along the other—that is completely absent in the pristine, uniform system. Ultimately, these findings suggest that extrinsic thermal templating could offer a powerful structural method to significantly enhance and reshape proximity-induced SOC, unlocking complex spin textures that set a robust foundation for future work. In particular, direct spectroscopic probes such as STS and ARPES measurements on rippled heterostructures would provide a natural next step toward experimentally mapping the predicted helical band structure and spin texture, opening further avenues for spintronic device design.

%\section*{Data Availability}
%The data that support the findings of this study are available within the paper and its Appendix. The datasets generated during and/or analyzed during the current study (specifically density functional theory input/output files) are available from the corresponding author on reasonable request.

%\section*{Code Availability}
%The density functional theory calculations were performed using the VASP package (https://www.vasp.at/). Post-processing and plotting were performed using pyProcar and VESTA. The code used for the Moiré ladder model and numerical diagonalization is available from the corresponding author on reasonable request.

%\section*{Competing Interests}
%The authors declare no competing interests.

%\section*{Author Contributions}
%F.M. and C.P. conceived the project and designed the experiments. 
%F.M. performed the theoretical modeling and DFT calculations. 
%M.F. performed DFT calculations. 
%V.G. synthesized the samples. 
%H.C.M. and C.Z.Z. contributed to the analysis of the STM data. 
%P.M. developed the Moiré ladder model and contributed to the theoretical interpretation. 
%F.M, C.P., P.M. and H.C.M wrote the manuscript with input from all authors.

\begin{acknowledgments}
C.P. acknowledges support from Fondecyt 1260378, Fondequip EQM200085, ANID CCTVal CIA250027, ANID SET CTI250019. F.M.  acknowledges support from Fondecyt grants 1231487 and 1260378, CEDENNA CIA250002 from ANID, and partial support by the supercomputing infrastructure of the NLHPC (CCSS210001). P.M. acknowledges support from the Fondo Nacional de Desarrollo Científico y Tecnológico (Fondecyt) under Grant No. 1250122. Work at Stanford (C.Z.Z. and H.C.M.) was supported by the US Department of Energy, Office of Science, Basic Energy Sciences, Materials Sciences and Engineering Division, under Contract DE-AC02-76SF00515. C.P. and H.C.M. thank the Luksic Scholars Foundation for fellowship support. 
\end{acknowledgments}

\appendix

\section{Appendixes}
\subsection{Experimental  Methods}
\label{sec:app-experimental}
TI/SLG nano-heterostructures based on thin Sb$_2$Te$_3$ nanoplatelets were grown on single-layer graphene. Chemical vapor deposition (CVD) growth of single-layer graphene on 25 $\mu$m copper foil ($>$99.8\% purity, Alfa Aesar) was carried out at 1050 °C using high-purity CH$_4$ and H$_2$ as precursor gases (20 sccm and 10 sccm, respectively). Growth pressure was kept at 0.18 mtorr. After growth, the SLG/copper sample was naturally cooled down in an Ar atmosphere \cite{Parra2017}. The resulting SLG on Cu samples was approximately 10 × 8 cm². Sections of $1\times1$ cm$^2$ were cut for the transfer of SLG onto SiO$_2$ substrates (n-doped, 285 nm oxide thickness) using a PMMA-assisted transfer method \cite{Parra2021}. Representative Raman spectrum is shown in Fig. \ref{fig:STM}c, which displays sharp G (1584 cm\textsuperscript{-1}) and 2D (2680 – 2693 cm\textsuperscript{-1}) bands and  I\textsubscript{2D}/I\textsubscript{G }$>$2, which is consistent with single-layer graphene.

The TI synthesis parameters were tuned to obtain nanostructures thinner than 30 nm in order to promote the presence of surface states and their interaction with the SLG substrate. Sb$_2$Te$_3$ NPs were synthesized by a catalyst-free vapor transport and deposition process. Antimony telluride powders (99.999\%, Alfa Aesar) were placed on a quartz plate in the center of one zone of the two-zone furnace (TFM2-1200 Across International) while the substrates were placed 11-15 cm downstream from the powder source. Additional variations of synthesis parameters were previously optimized in a separate study \cite{Gallardo}. Sb$_2$Te$_3$ NPs growth was carried out for 5 min at a temperature of 500 °C and with an Ar flow of 50 sccm. Once the growth is complete, the system was naturally cooled down ($\sim3.5^\circ$C/min), maintaining the Ar flow and a 0.3 torr system pressure. The resulting nano-heterostructures were characterized by Raman, AFM, and low-temperature scanning tunneling microscopy (LT-STM) to obtain information on their compositional and structural properties. 
The representative Raman spectrum of the Sb$_2$Te$_3$ NPs (Fig. \ref{fig:STM}d) exhibits three characteristic peaks located at 69.1 cm\textsuperscript{-1}, 115.4 cm\textsuperscript{-1}, and 164.7 cm\textsuperscript{-1}. The peak at 115.4 cm\textsuperscript{-1} is associated with the in-plane vibrational mode E$^2$$_g$ of rhombohedral Sb$_2$Te$_3$, whereas the peaks at 69.1 cm\textsuperscript{-1} and 164.7 cm\textsuperscript{-1} correspond to the out-of-plane vibrational modes A$^1$$_1$$_g$ and A$^2$$_1$$_g$ , respectively. These Raman features are consistent with previously reported spectra of Sb$_2$Te$_3$, confirming the formation of single-crystalline nanoplatelets synthesized by CVD. 

These experimentally resolved structural features provide the basis for the theoretical analysis presented in this work, which explores the electronic consequences of the observed nanoscale corrugation.

\subsection{Computational Methods}
\label{sec:computational}

We employed different methods to achieve the best balance between accuracy and computational burden, depending on the specific goal (see below). In general, we used the VASP software \cite{vasp1,vasp2,vasp3,vasp4} and PAW pseudopotentials \cite{paw}.

In Sec.~\ref{sec:theo_origin}, to get accurate energetics and energies, we used the functional SCAN+rVV10\cite{rvv10}, which provides an accurate description of van der Waals interactions. The kinetic energy cutoff was set to 600 eV, and a $21\times21$ k-point grid was used. We tested the effects of spin-orbit coupling (SOC) and found it to be negligible for relative energies. 

However, in Sec.~\ref{sec:electronic}, the attention is shifted to the electronic structure, and then we used a lower accuracy setting as it has been well documented that the Perdew-Ernzenhof-Burke (PBE) \cite{PBE} correctly describes the low-energy behavior together with a van der Waals correction\cite{DFT-D3}. The kinetic energy cutoff was lowered to 500 eV. The k-point mesh was $5\times3\times1$ for the orthogonal supercell and decreased along the long direction when using a supercell.

For the analysis and plotting, we used the pyProcar \cite{pyprocar,pyprocar2} and Vesta software \cite{vesta}. 

To simulate the buckling, we modulated the out-of-plane positions ($z$) of the heterostructure by $z=h\cos(2\pi y/L)$. With $h=2.5$\AA{}, and $L$ being the ripples wavelength ($\sim 8.7$~nm). The closest value was achieved using a $ 1\times 12$ supercell. 

To capture the shrinking of the unit cell due to the ripples, we changed the corresponding lattice vector to keep the correct distances of the flat graphene layer. The arc length of the graphene layer is
\begin{equation}
    L = \int_{0}^{L} \sqrt{1 + \left[\frac{2\pi a}{L}\cos\left(\frac{2\pi x}{L}\right)\right]^2} \, dx
\end{equation}
Numerical evaluation gives $\frac{L_0}{L}\approx 1.007$, \textit{i.e.} the unit cell should shrink less than 1\% to avoid extra strain due the ripples.

\subsection{Heterostructure without ripples}
\label{sec:flat}

The defining feature of Moiré superstructures is their potential to host exotic electronic states. To isolate the theoretical electronic effects that ripples of the type observed experimentally could induce, we first establish a theoretical baseline by calculating the electronic structure of an ideal, flat $\text{Sb}_2\text{Te}_3$/graphene heterostructure using DFT (see Appendix for details). This reference system consists of a $1\times1$ unit cell of Sb$_2$Te$_3$ on a $\sqrt{3}\times\sqrt{3}$ graphene supercell, with a minimal lattice mismatch of 0.35\% (see Fig.~\ref{fig:geometries}b). In its bulk form, Sb$_2$Te$_3$ is a well-known 3D TI, characterized by a metallic surface state protected within its bulk band gap of $E_{gap} \sim0.19$ eV~\cite{Zhang2009, Mohelsky2024}.

\begin{figure}[ht]
    \centering
    \includegraphics[width=\columnwidth]{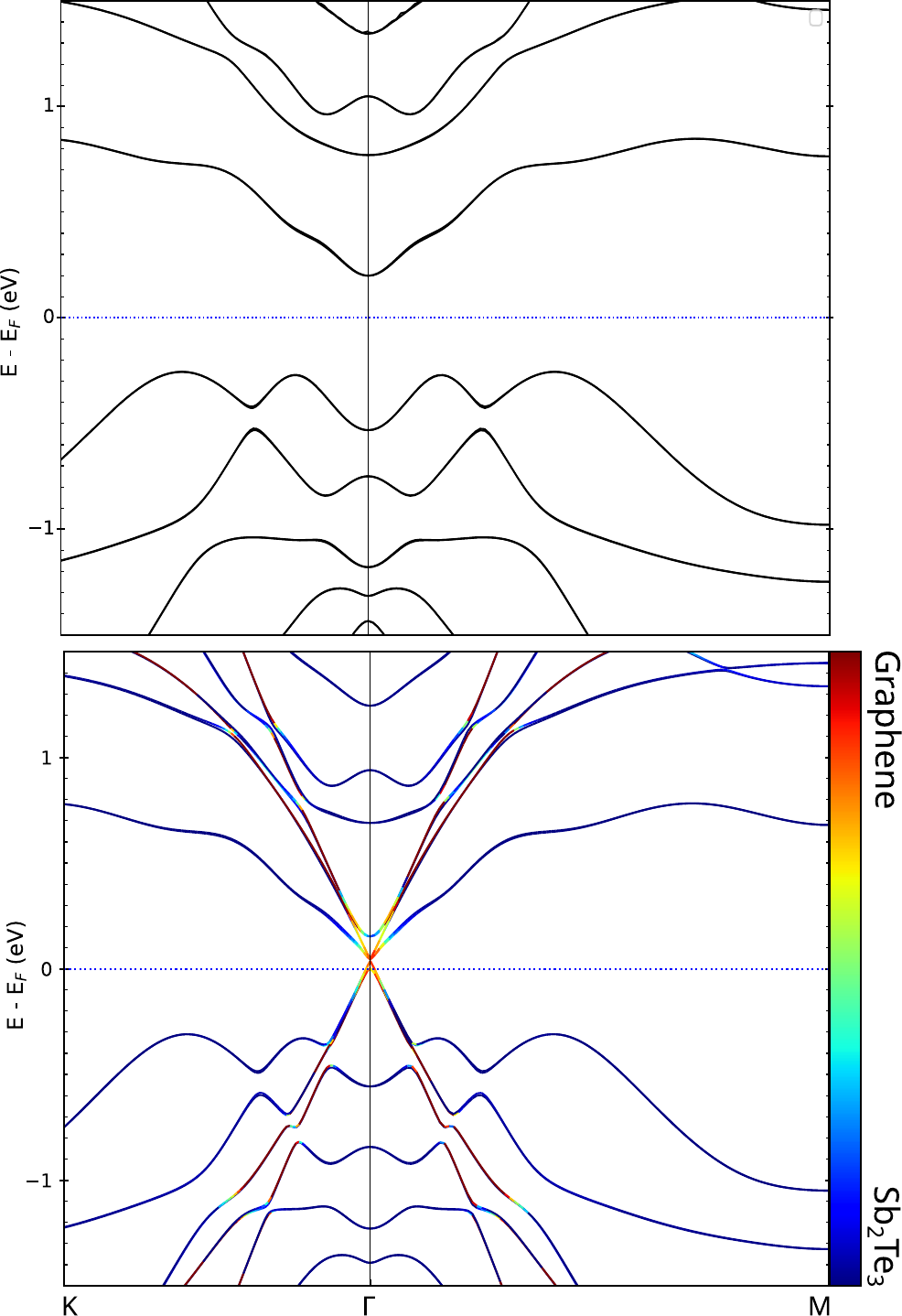}
    \caption{Band structure of a single QL of Sb$_2$Te$_3$: freestanding (top) and on graphene (bottom). SOC is included. The Dirac cone from graphene is folded to $\Gamma$. At this energy scale, the bands appear spin-degenerate; the Rashba spin texture becomes visible only at much higher magnification.}
    \label{fig:1x1-bands}
\end{figure}

Figure~\ref{fig:1x1-bands} (top panel) shows the band structure of a freestanding single QL (\textit{i.e.} without graphene). It reveals a gap of $0.45$ eV, arising from the hybridization between the upper and lower surface states. This gap signifies the loss of topological character in the ultrathin limit, caused by wavefunction overlap between the slab's surfaces. Such quantum-confinement effects are well documented in thin-film TIs; for instance, the related Bi$_2$Se$_3$ is experimentally confirmed to become fully insulating below six QLs~\cite{Zhang2010}. %For comparison, excluding spin-orbit coupling (SOC) yields a larger trivial gap of $0.73$ eV.

When the Sb$_2$Te$_3$ thin film is placed on graphene (Fig.~\ref{fig:1x1-bands}, bottom panel), there is a strong hybridization of the bands at $\Gamma$, right at the Fermi level. This is a direct consequence of the $\sqrt{3}\times\sqrt{3}$ reconstruction, which folds the graphene K and K' valleys to the zone center. At $\Gamma$, the graphene contributes eight bands (2 valleys $\times$ 2 spins $\times$ 2 branches) that become nearly degenerate with the Sb$_2$Te$_3$ states. 

A zoom into the band structure reveals that this interaction lifts the degeneracy, splitting the bands into two massive Dirac cones with gaps of 10 and 40 meV. This behavior can be understood as a \textit{Dirac-Dirac resonance}: the folding forces the graphene and TI Dirac cones to align in both momentum ($\Gamma$) and energy. The resulting hybridization is strong enough to open a global gap in the heterostructure, effectively rendering the TI surface states sensitive to the graphene layer—a surprising deviation from the behavior of thicker TI films, where the surface states typically remain robust. These results agree with earlier calculations \cite{Jin2013} and measurements \cite{Bian_2016}. In other typical TI materials, a larger mismatch shifts both Dirac cones in energy, avoiding this resonance \cite{Naimer2023,song2018spin}.

For the sake of comparison, if SOC is not included, there is no such strong hybridization at $\Gamma$; the bands resemble the top panel of Fig.~\ref{fig:1x1-bands} with the graphene Dirac cone superimposed, along with minor avoided crossings.

In thicker Sb$_2$Te$_3$ samples on graphene, this strong Dirac-Dirac resonance is diminished. As the top and bottom surface states decouple, only the bottom surface state physically interacts with the substrate. Consequently, there is only one degenerate Kramers pair at $\Gamma$ available to interact with the graphene's folded Dirac cones (which constitute two Kramers pairs), leading to a fundamentally different and less effective hybridization landscape \cite{zhang2014proximity,Lin2017}.

\subsection{Energy considerations}
\label{app:simpleEnergy}
\begin{figure}
    \centering
    \includegraphics[width=\columnwidth]{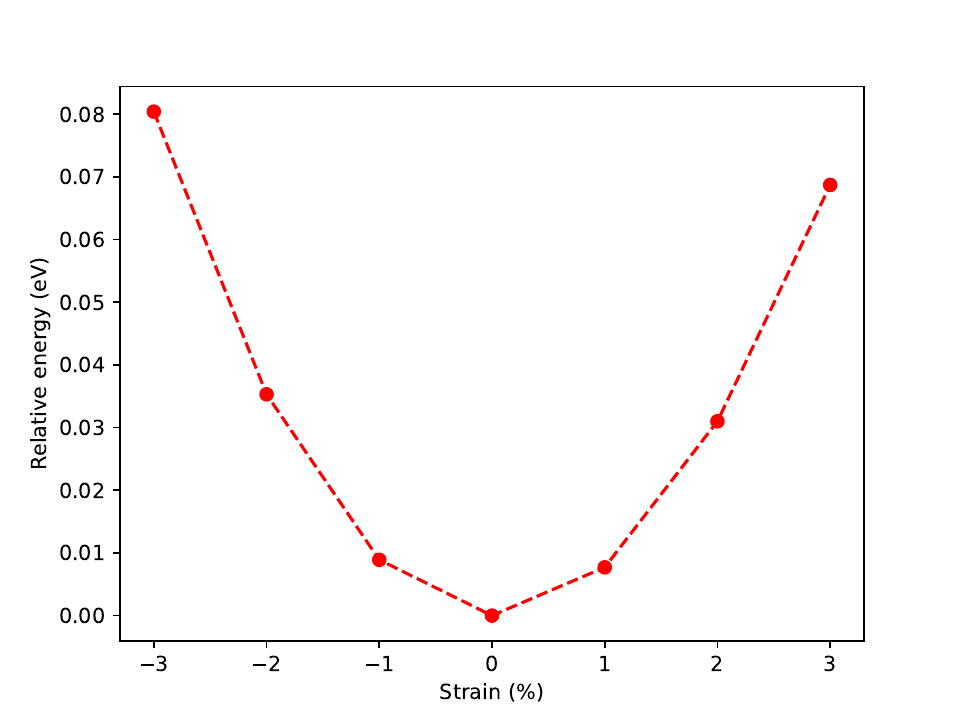}
    \caption{Energy of a single QL of Sb$_2$Te$_3$ as a function of biaxial strain. The energy of the minimum energy system is set to zero.} 
    \label{fig:strainstress}
\end{figure}

According to DFT calculations, the distance between graphene and Sb$_2$Te$_3$ is 3.59 \AA, which is typical for van der Waals (vdW) heterostructures. The measured samples preserve the periodicity of the unit cell along one direction; this aspect is not considered further in the present discussion. In the orthogonal direction, the samples exhibit a periodicity of $L\approx 8.7$ nm, corresponding to approximately 20 or 21 unit cells of Sb$_2$Te$_3$.
The binding energy ($E_{B}$)  between graphene and single Sb$_2$Te$_3$ QL, 
\(E_{B} = E_{total} - E_{graphene} - E_{Sb_2Te_3}=0.23\) eV per cell (\textit{i.e.} the cell from Fig.~\ref{fig:geometries}b). The variation in $E_B$ resulting from a lateral shift between the two materials is less than 0.01 eV per cell. Consequently, in the absence of point defects, the energy required to slide the layers relative to each other is negligible compared to thermal energy at room temperature. The change in energy of the Sb$_2$Te$_3$ monolayer under biaxial strain, as shown in Fig.~\ref{fig:strainstress}, is also negligible (0.01 eV per unit cell) for a strain of 1\%. Uniaxial strain is expected to result in even lower stress. Since the energy penalty for sliding is less than 0.01 eV, and both energy penalties are an order of magnitude smaller than $E_B$, significant strain accumulation is unlikely.

In a perfectly homogeneous material, the total strain energy is $\sim30$ meV, corresponding to each unit cell being subjected to a $0.35$\% strain. Since the system is free to slide, the actual strain energy may be lower. This value is negligible compared with the binding energy ($E_B=0.23$ eV per unit cell). To further evaluate the energy cost associated with potential buckling, we consider the buckling amplitude $ z(y) = \alpha\cos\left(\frac{2\pi y}{L}\right)$
Here, $2\alpha\approx 5.0$~\AA~ represents the observed height difference. The variable $y$ denotes the direction of the stripes. The minimum curvature radius ($R$) occurs at $y=0$, and the corresponding curvature ($\kappa=R^{-1}$) $\kappa = \frac{|z^{''}|}{(1+z^{'2})^{3/2}}\sim 0.013$ \AA$^{-1}$. 
The bending energy per atom, $E_{bend} = \frac{C_bk^2}{2}$, is determined by the material's bending modulus. For graphene, $C_b=3.43$ eV/\AA$^2$ per atom \cite{gonzalez2018}, which yields a total bending energy of $E_{bend}\approx0.02$ eV per graphene strip. While Sb$_2$Te$_3$ quintuple layers exhibit anisotropy, a reasonable estimate can be made by applying the bending energy expression for homogeneous materials, $E_{bend} = \frac{Yt^2\kappa^2}{2}$, where $Y$ denotes Young's modulus and $t$ represents the material thickness. The $t^2$ dependence indicates that bending Sb$_2$Te$_3$ is considerably more challenging than bending graphene. The bending modulus of the transition metal dichalcogenide MoS$_2$ is approximately ten times greater than that of graphene ($C_b\approx45$ eV/\AA$^2$ per atom \cite{gonzalez2018}). Although Sb$_2$Te$_3$ and MoS$_2$ differ in several properties, this value provides a reasonable approximation because Sb$_2$Te$_3$ is more than twice as thick as MoS$_2$, and in nanofilm experiments, the Young's modulus of Sb$_2$Te$_3$ is about five times smaller than that of MoS$_2$. These two factors are expected to compensate for each other, leading to an estimated bending energy of $E_{bend}=0.2$ eV per stripe.

The bending energy is expected to be similar to the stress induced by the mismatch. Disregarding any relative sliding, a buckling pattern should induce both compression and elongation, overall increasing the stress. Therefore, the observed ripples are not caused by an intrinsic instability of the freestanding Graphene/Sb$_2$Te$_3$ heterostructure. On the other hand, the bending energy at the minimum curvature radius is similar to the binding energy, $E_B-E_{bend}\ge0$, thus separation of the two two-dimensional materials is ruled out; if buckling occurs, both subsystems remain together. Additionally, Sb$_2$Te$_3$ samples thicker than 1QL are expected to either separate from graphene or flatten it, as $E_{bend}$ increases quadratically with the sample thickness. 

An alternative explanation for stripe formation is that the system exhibits thickness variations along the stripes.  These variations could result in distinct material phases. Lateral coexistence of different phases requires that the process be energetically favorable. This requirement implies that the binding energy between graphene and the Sb$_2$Te$_3$ QL should increase across regions of varying thickness. The system must also compensate for the energy cost associated with lateral defects or interface reconstructions between the two SbTe-based phases. The unknown phase is likely strained by the graphene layer, which is necessary to explain the observed transition between the two phases; otherwise, only the lower-energy phase would be present. The formation of stripes supports the existence of a periodic strain field. Furthermore, an energy barrier must be present to prevent sliding, as its absence would inhibit the accumulation of strain. This hypothesis is also applicable to other materials. Several structurally distinct phases are observed for chalcogens and pnictogens, each characterized by unique atomic arrangements and stoichiometries. The most prevalent among these are layered structures containing interwoven Bi$_2$ interlayers \cite{Kim2001,Bos2007}. These phases can coexist within a single sample \cite{drzewowska2023different}. Thinner phases of As$_2$Se$_3$, distinguished by more complex geometries, have also been identified \cite{LITYAGINA2015}. Additionally, unusual forms, such as a square BiTe monolayer phase, have been reported on Bi$_2$Te$_3$ \cite{wang2017}. Of particular relevance is a high-pressure-induced rectangular crystal phase of BiS$_2$ and BiSe$_2$ \cite{Yamamoto2015,Kevy2019}. The freestanding monolayer of SbTe$_2$, referred to as r-SbTe$_2$, is predicted to be stable in its distorted rectangular geometry \cite{Mella2024}. However, a simple analysis shows that the combined energetic penalties from higher formation energy and increased phase-boundary costs are challenging to overcome with a modest binding-energy enhancement in our case.
This conclusion is further supported by two additional considerations related to phase behavior. First, the hypothesis that stripes arise from different Sb$_x$Te$_{1-x}$ phases separated by boundaries would require those boundaries to be pinned by a substantial energy barrier, which is at odds with our earlier finding of minimal sliding resistance between layers. Second, the gently undulating features observed by STM suggest mechanical buckling, as opposed to the sharp, step-like transitions expected at lateral phase boundaries between distinct Sb$_x$Te$_{1-x}$ phases. Together, these observations rule out the coexistence of distinct Sb$_x$Te$_{1-x}$ phases as a likely explanation for the stripes.

For the sake of completeness, long molecular dynamics simulations were conducted using a DFT-based force field for various arrangements and strain values on the Sb$_2$Te$_3$ QL. The system was allowed to evolve freely, subject only to periodic boundary conditions. In all cases, neither bending, buckling, nor stripe formation was observed.

\subsection{Moiré ladder model for modulated Sb$_2$Te$_3$/SLG}
\label{app:effective}
\begin{figure*}
\centering
\includegraphics[width=\linewidth]{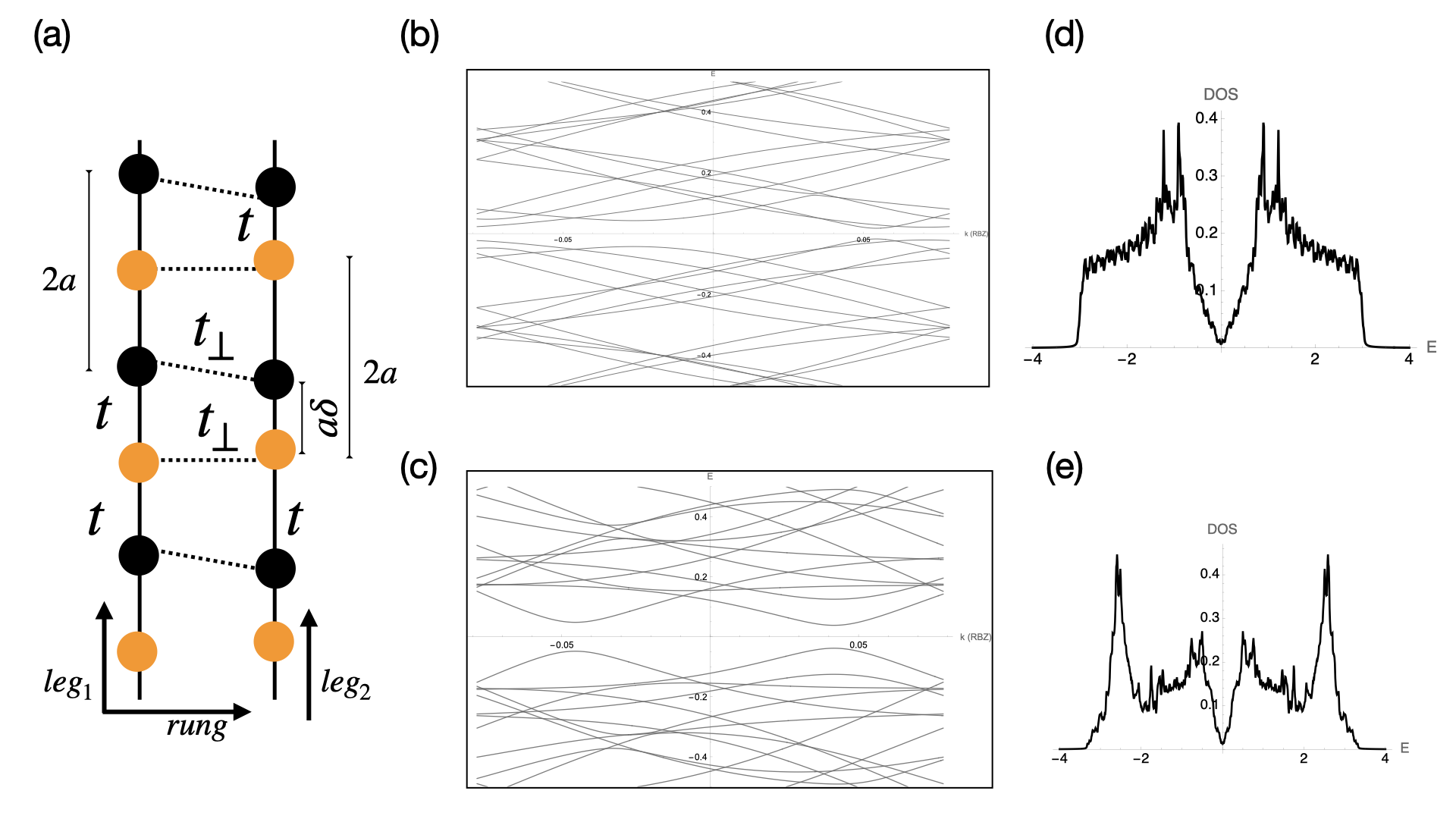}
\caption{\label{fig:model1}(a) The moiré ladder model. (b,c) Spectrum of the minibands of the ladder model in the RBZ with $\delta=\frac{17}{20}$ and $\alpha=t$ without (b) and with (c) spin orbit coupling. (d) DOS at T=0 of a spinless system with $\delta=\frac{19}{20}$ and (e) a spinful system with spin orbit coupling, $\alpha=t$ and at $\delta=\frac{19}{20}$.}
\label{fig:model_1}
\end{figure*}
\begin{figure*}
\centering
\includegraphics[width=\linewidth]{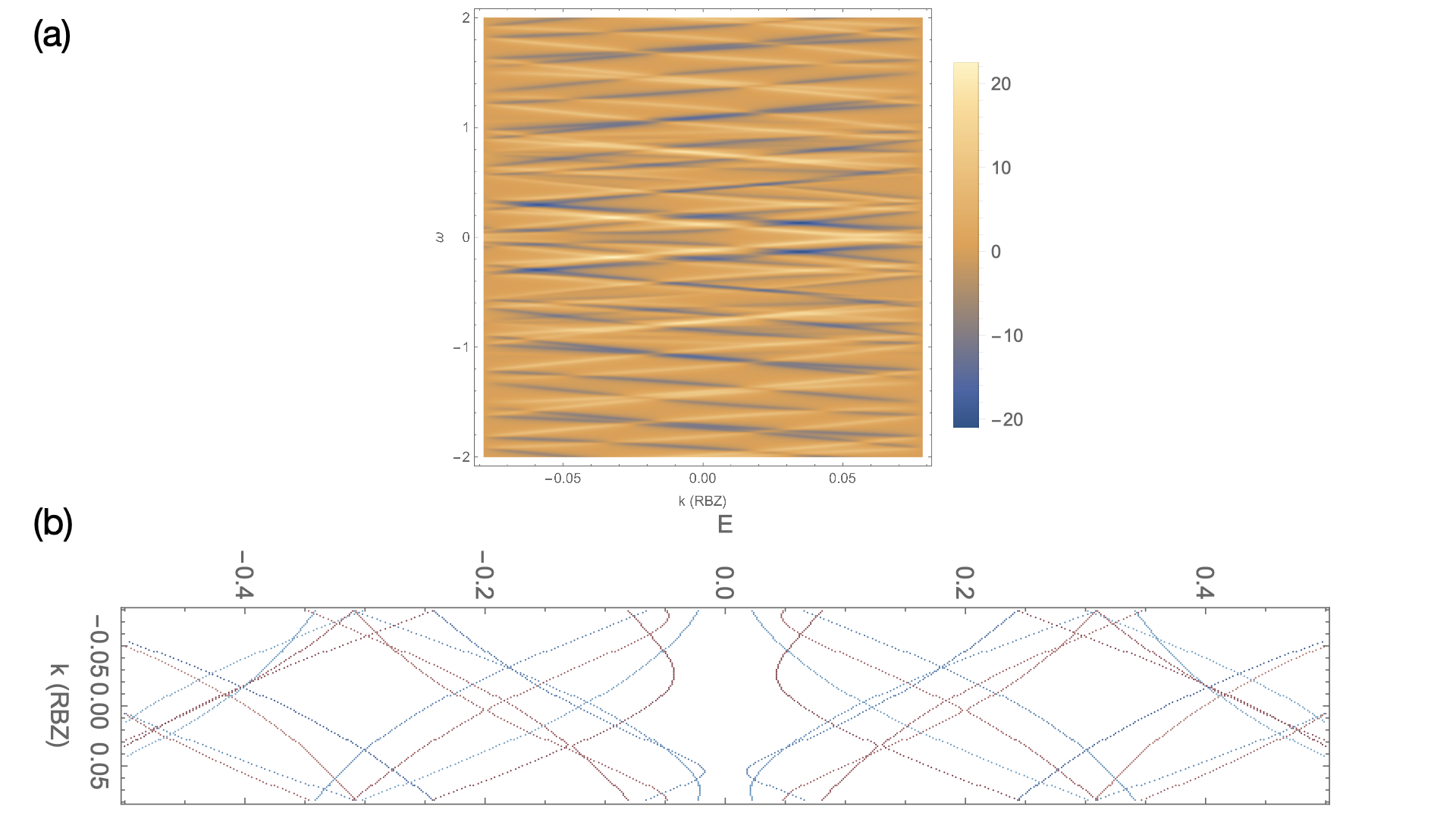}
\caption{\label{fig:model2}(a) Helical spectral function of the ladder model at $\delta=\frac{19}{20}$, $\alpha=t$ and $T=0$. (b) zoom out of the minibands spectrum close to the fermi energy (E=0) at $\delta=\frac{19}{20}$ and $\alpha=t$. The colors of the minibands reflect their helicity weight.}
\label{fig:model_2}
\end{figure*}
In this appendix, we analyze a simplified model to assess the impact of ripple-induced moiré modulation on the spectrum of a spin–orbit-coupled heterostructure. 
We model Sb$_2$Te$_3$1QL/SLG heterostructures as a ladder system, where the top and bottom legs represent the top and bottom surfaces of the heterostructure, as illustrated in Fig.\ref{fig:model1}(a). The ladder consists of two coupled legs, each with two sites (A and B representing Sb and Te atoms) per unit cell. The composite ladder unit cell contains a total of four sites $\{A_1,B_1,A_2,B_2\}$ (subscripts $1,2$ label legs).  The composite cell has a length equal to $2a$, and the original reciprocal period is \(
G_0=\pi/a\). In our effective description, the structural ripples of the Sb$_2$Te$_3$ film enter through a long-wavelength modulation of the interleg hopping and a slight dimerization in leg two.  In real space, this modulation is \(t_{\perp}(n)\sim\cos(2\pi\delta\, n),\) where $n$ is the site index along the ladder direction and $\delta=p/q\in (0,1)$ is a positive dimensionless parameter that determines the period of the modulation.  In addition, $\delta\sim 1$ controls the weak dimerization in leg~2 which gives rise to a Peierls phase in this leg.  This choice reflects the fact that the ripple profile modulates the local bonding environment of leg~2 of the model (representing the interface-facing side of the quintuple layer), while leg~1 remains undimerized. Intraleg hopping occurs between nearest-neighbor atoms, and its amplitude is $t$.

To relate $\delta$ to the physical ripple wavelength
$\lambda_{\mathrm{ripple}}$, we match the modulation wavevector to the
ripple wavevector $Q_{\mathrm{ripple}}=2\pi/\lambda_{\mathrm{ripple}}$.
Since the dimensionless ratio of the two wavevectors is
\(
\delta = \frac{Q_{\mathrm{ripple}}}{G_0}
        = \frac{\,2\pi/\lambda_{\mathrm{ripple}}\,}{\pi/a}
        = \frac{2a}{\lambda_{\mathrm{ripple}}},
\)
it follows that
\(
\lambda_{\mathrm{ripple}} = \frac{2a}{\delta}.
\)
To render the modulation commensurate, we choose $\delta=p/q$ with $p$ and $q$ coprime. Interleg modulation and the dimerization of leg-2 are periodic after $q$ unit cells, so that
one ripple period spans $q$ unit cells and has physical length
\(
\lambda_{\mathrm{ripple}}= \frac{2a\,q}{p}.
\)
For instance, for $\delta=0.95$ ($p=19$, $q=20$), the resulting
supercell contains $20$ unit cells and the ripple wavelength is
$\lambda_{\mathrm{ripple}}\simeq 40a$, consistent with the mesoscopic ($\sim$5--20\,nm) corrugation length scales observed experimentally in Sb$_2$Te$_3$/graphene heterostructures. Henceforth, we choose \textit{a}, half the legs lattice constant, as the length unit.

The relative mismatch between legs yields a moiré supercell of \(q\) composite cells of \(4q\) sites, a moiré reciprocal vector in the reduced Brillouin zone (RBZ) \(b_s=\frac{G_0}{q}=\frac{\pi}{q}\), and a real-space supercell of length \(L_s=2\,q\). Rung tunneling occurs between the same sublattice, and its average is \(t\). Schrieffer–Wolff perturbation theory \cite{schrieffer1966relation} gives rise to a first-harmonic of amplitude \(t_1=(1-\delta)t\) modulated by the envelope $t_\perp(n)=t+t_1\cos\left(\frac{2\pi n}{q}\right)$. Throughout, we choose \(t\) as our energy scale. 

Structure factors associated with hopping in leg-1 and leg-2 are  respectively \(C(k)=2\cos k\), \(C_\delta(k)=2\cos\!\big(k-\pi\delta\big)\), and the Peierls phase from the dimerization of leg-2 is $\phi(k)=k(1-\delta)+\pi\delta$. In Fourier space, the moiré rung modulation becomes \(t_\perp(k)=1+(1-\delta)\cos\!\big(qk\big)\). Throughout, we choose \(t\) as our energy scale.

In graphene-based heterostructures coupled to a topological insulator, a perpendicular electric field is induced, and the dominant proximity-induced SOC is generally of Rashba character \cite{bihlmayer2022rashba,ado2017microscopic,chen2018electric}. In a quasi-one-dimensional lattice description, the Rashba SOC generated by a perpendicular electric field $\mathbf{E}=E_z\hat{z}$ takes the form
\begin{equation}
H_{\rm SOC}
=
\alpha \sum_j 
\left(
c^\dagger_{j+1}\, i\sigma_y\, c_j
-
c^\dagger_{j}\, i\sigma_y\, c_{j+1}
\right),
\label{eq:soc_realspace}
\end{equation}
where the fermion operators $c_j=(c_{j\uparrow},c_{j\downarrow})^T$ and $\sigma_y$ act on spin. Fourier transforming Eq.~\eqref{eq:soc_realspace},
one obtains
\begin{equation}
H_{\rm SOC}^{(1)}
=
2i\alpha \sum_k
 \left(\sin k \right)
c_k^\dagger \sigma_y c_k .
\label{eq:soc_kspace}
\end{equation}
In a uniform system such as leg-1, the SOC term in Eq.\ref{eq:soc_kspace} appears as a momentum-odd contribution, which vanishes at time-reversal invariant momenta and reflects the inversion-symmetry breaking at the interface. In leg-2 the the hopping paths of electrons are modulated in real space, and as a result, the SOC term on that leg acquires a phase,
\begin{equation}
H_{\rm SOC}^{(2)}
=
\alpha \sum_j 
\left(
e^{-i\pi\delta}\, c^\dagger_{j+1,2}\, i\sigma_y\, c_{j,2}
-
e^{+i\pi\delta}\, c^\dagger_{j,2}\, i\sigma_y\, c_{j+1,2}
\right),
\label{eq:soc_dimer_real}
\nonumber
\end{equation}
Fourier transforming the previous equation yields
\begin{align}
H_{\rm SOC}^{(2)}
&=
2 i \alpha\sum_k
 e^{i\phi(k)}\sin(k-\pi\delta)\;
c^\dagger_{k,2}\,\sigma_y\, c_{k,2}.
\label{eq:soc_dimer_k}
\end{align}
with $\alpha$ the SOC strength and
\(\sigma_y 
\) the second Pauli matrix.
We define the spinor as the eight-component object,
\(
\Psi_{k}=
(c_{1A\uparrow},c_{1,A\downarrow}, c_{1,B\uparrow},c_{1,B\downarrow},c_{2,A\uparrow},c_{2,A\downarrow},c_{2,B,\uparrow},c_{2,B\downarrow})\)
and  we assign opposite helicity to the two legs, such that leg~1 carries helicity $+\sigma_y$ (upper surface) and leg~2 carries helicity $-\sigma_y$ (lower surface). The total Hamiltonian, including both the kinetic terms and the opposite-helicity spin–orbit coupling, yields 
\begin{widetext}
\begin{equation}
\label{eq:H}
H(k)=
\begin{pmatrix}
0 & 0 & C(k) & -iS(k)& t_\perp(k) &0&0 &0 \\
0 & 0 & iS(k) & C(k)&0&t_\perp(k)&0 &0\\
C(k) & -iS(k)&0 & 0&0 & 0& t_\perp(k) &0\\
iS(k) & C(k)&0 & 0&0 & 0&0&t_\perp(k)\\
t_\perp(k) & 0 & 0 & 0&0 & 0&C_\delta(k)\,e^{i\phi(k)}&iS_\delta(k)\,e^{i\phi(k)}\\
0 & t_\perp(k) & 0 & 0 &0& 0&-iS_\delta(k)\,e^{-i\phi(k)}& C_\delta(k)\,e^{i\phi(k)}\\
0&0&t_\perp(k)&0&C_\delta(k)\,e^{-i\phi(k)}&iS_\delta(k)\,e^{i\phi(k)}&0&0\\
0&0&0& t_\perp(k)&-iS_\delta(k)\,e^{-i\phi(k)}&C_\delta(k)\,e^{-i\phi(k)}&0&0\\
\end{pmatrix},
\end{equation}
\end{widetext}
where $S(k)=2\alpha \sin k$ and $S_\delta(k)=2 \alpha \sin(k-\pi\delta)$.  Eq.\ref{eq:H} is given in the extended Brillouin zone (EBZ), \(\rm k\in (-qG_0,qG_0 )\), with reciprocal superlattice 
\(G_m = \{ mqG_0\ |\ m\in\mathbb{Z}\}\). For a given $\delta$, the similarity transformation $\Delta=\pi q$  applied to $H(k)$ imposes constraints on p and q that determine the generic global period $2\pi q$ giving rise to EBZ \cite{mellado2025sliding}. In the absence of SOC, the spectrum obeys $E(k)=E(k+G_m)$. Once SOC is introduced, the Hamiltonian acquires momentum-odd contributions $\propto \sin(k-\pi\delta)\,\sigma_y$, which reverse sign under $k\rightarrow k+G_m$. Consequently, translation by $G_m$ alone ceases to be a symmetry, and the Hamiltonian remains invariant under the combined action of a moiré translation and a $\pi$ spin rotation about the $y$ axis. Because this composite symmetry squares to a translation by $2G_m$, the spectral periodicity in the extended Brillouin zone is effectively halved.
 
The reduced first Brillouin zone of the moiré lattice (RBZ) spans the interval \((-b_s/2,b_s/2)\), and the reciprocal superlattice 
\(G_s = \{ m\,b_s\ |\ m\in\mathbb{Z}\}\). To build the Hamiltonian of the system in the RBZ we keep the \(q\) replica momenta \(\{\bar{k}+m G_s\}_{m=0}^{q-1}\) in the Hamiltonian matrix $H_{\mathrm{RBZ}}$. Thus, the block‐diagonal piece becomes
\(
\big[H^{(0)}_{\mathrm{RBZ}}(\bar{k})\big]_{m m'}=
\delta_{m m'}\,H\!\big(\bar{k}+m G_s\big),\) 
\( m,m'=0,\dots,q-1
\). Rung modulation \(\cos(qk)\) connects nearest-neighbor replicas in the RBZ and gives rise to the off-diagonal part \(\big[\delta H_{\mathrm{RBZ}}(\bar{k})\big]_{m,m\pm 1}=\frac{t_1}{2}\,R,\quad
R = \sigma_1 \otimes \mathbb{I}_2\), where $\sigma_j$ is the j-$th$ Pauli matrix. The  \(4q\times4q\) moiré Hamiltonian in the RBZ becomes \cite{mellado2025sliding}
\(H_{\mathrm{RBZ}}(\bar{k})=H^{(0)}_{\mathrm{RBZ}}(\bar{k})+\delta H_{\mathrm{RBZ}}(\bar{k})\).
If \(t_1=0\) the \(4q\) bands split into \(q\) independent copies of the four parent bands. The symmetry of \(H\) is preserved by $ H_{\mathrm{RBZ}}$ \cite{mellado2025sliding}. 
Fig.\ref{fig:model1}(b,c) shows a zoomed-out view of the energy spectra of the minibands in the reduced Brillouin zone at $\delta=0.95$ and $\delta=0.85$ in the absence and presence of spin-orbit coupling. Without SOC, the spectrum consists of multiple folded minibands that intersect extensively, reflecting both the moiré backfolding and the underlying spin degeneracy. Fig.\ref{fig:model1}(c) shows that SOC lifts spin degeneracy, and a large fraction of the band crossings is replaced by avoided crossings. This reconstruction reflects the entanglement of spin and moiré degrees of freedom induced by SOC and leads to a redistribution of spectral weight across the miniband manifold as predicted by our DFT calculations presented above. With a perfect match between the legs ($\delta=1$), the density of states depicts a pair of van Hove singularities (VHS) at the edges of the bands, a finite density of states at all energies, and another pair of VHS at $ E=\pm t_1$, signaling the coupling between the two legs \cite{mellado2025sliding}. Once $\delta\neq 1$, the moiré potential smears out the peaks at the edges of the spectra; the VHS at $E=\pm t$ splits into two peaks about $E\sim\pm t_1$  due to the breaking of inversion symmetry Fig.\ref{fig:model1}(d) \cite{ribeiro2015origin}.  Including SOC, Fig.\ref{fig:model1}(e) removes spin degeneracy, opening gaps at many band crossings, redistributing spectral weight by increasing the population of states close to the Fermi level and producing a qualitatively altered DOS. 

The Dirac-like crossings observed in Fig.\ref{fig:model1}(c) indicate that the moiré-modulated SOC not only lifts degeneracies but also reorganizes the miniband structure into effective two-band sectors with linear dispersion. The case at $\delta=\frac{17}{20}$ illustrates how moiré modulation and SOC jointly induce emergent relativistic behavior in otherwise heavily folded miniband spectra. With SOC present, a pure moiré translation by a reciprocal lattice vector $G_m$ ceases to be a symmetry. Nevertheless, the Hamiltonian is invariant under the composite operation $\tilde T = T_{G_m}\mathcal R_y(\pi)$, with $\mathcal R_y(\pi)=-i\sigma_y$ denoting a $\pi$ spin rotation about the SOC-defined axis.  $\tilde T$ acts as a quantum number that differentiates miniband eigenstates, such that bands with distinct $\tilde T$ eigenvalues cannot hybridize at the same momentum. As a result, when moiré folding brings bands from different $\tilde T$ sectors into contact, SOC imposes symmetry-protected linear crossings, yielding Dirac-like nodes in the miniband spectrum \cite{mellado2026moire}.

%Helical states have been observed in moiré heterostructures featuring proximity induced SOC \cite{takagaki2016manipulation,khokhriakov2020gate,kiemle2022gate,song2018spin,li2012topological,hoque2024room,tanabe2025high,zollner2024proximity}. In the case of Sb$_2$Te$_3$ , combined ARPES and electrical detection experiments provide strong experimental evidence for the existence and manipulability of the helical surface states where electron spin is locked perpendicular to momentum due to strong spin-orbit coupling and time-reversal symmetry \cite{li2016electrical,takagaki2016manipulation,takagaki2012robust,zollner2021heterostructures,yin2022moire,zhao2024layer,locatelli2022magnetotransport,manchon2015new}. 

Consider the helicity operator as the difference between the spin densities on the two legs \cite{bychkov1984oscillatory},
\(
\Gamma_{\mathrm{hel}}
=
S_y^{(1)} - S_y^{(2)},
\)
where
\(
S_y^{(\ell)}
=
\sum_k c_{k,\ell}^\dagger \, \sigma_y \, c_{k,\ell},
\)
with $\sigma_y$ acting in spin space. $\Gamma_{\mathrm{hel}}$ quantifies how unevenly spin polarization along the SOC-determined axis is shared between the two legs. It is antisymmetric under leg exchange, invariant under time reversal, and zero when SOC is absent due to spin-rotation symmetry.

To find out whether helical states are plausible in graphene/Sb$_2$Te$_3$ thin films, at the single-particle level, we compute the dynamical spectral function using Lorentzian broadening:
\(
A_\Gamma(k,\omega) = \sum_n
|\bra{n,k}{\Gamma}\ket{n,k}|^2
\frac{\eta}{(\omega - E_{n,k})^2 + \eta^2}
\), where $\eta$ is a small positive broadening. The helicity-resolved spectral function offers a direct probe of how proximity-induced spin–orbit coupling is redistributed throughout the moiré minibands. Without spin–orbit coupling, spin-rotation symmetry enforces the helicity-resolved spectral function to be zero. Upon introducing SOC, Fig.\ref{fig:model2}(a) reveals a dense mesh of oblique streaks with alternating blue and white intensities, spanning the reduced Brillouin zone. Each streak represents a miniband branch with nonzero helicity weight, where the color encodes both the sign and magnitude of the expectation value of the helicity operator $\Gamma_{\mathrm{hel}}$. The tilted alignment of these features arises from the finite group velocity of the associated states, and the alternating sign directly evidences spin-momentum locking: minibands with opposite dispersion slopes possess opposite helicity, as enforced by time-reversal symmetry.

Unlike uniform Rashba systems, where helicity is restricted to a few well-defined branches, the model shows that the moiré potential distributes helicity across numerous minibands. The color-coded miniband structure in the RBZ in Fig.\ref{fig:model2}(b) shows that every miniband branch in the reduced Brillouin zone carries a nonzero helicity weight. The simultaneous presence of positive and negative helicity on different branches reflects the odd-momentum character of the SOC and the maintenance of time-reversal symmetry. Fig.\ref{fig:model2}b shows that fluctuations in the helicity order parameter can be derived from an extensive set of particle-hole excitations involving all minibands, making the system intrinsically prone to helical collective behavior.

\bibliography{apssamp}% Produces the bibliography via BibTeX.

\end{document}